\documentclass{PoS}

\title{Nucleon structure from 2+1-flavor dynamical DWF ensembles}

\ShortTitle{DWF Nucleon}

\author{Michael Abramczyk\\
        Department of Physics, University of Connecticut, Storrs, CT 06269, USA\\
        E-mail: \email{michael.abramczyk@uconn.edu}
        }

\author{Meifeng Lin\\
        Computational Science Initiative, Brookhaven National Laboratory, Upton, NY 11973, USA\\
        E-mail: \email{mlin@bnl.gov}
        }
        
\author{Andrew Lytle\\
		School of Physics and Astronomy, University of Glasgow, Glasgow G12 8QQ, UK\\
		Email: \email{Andrew.Lytle@glasgow.ac.uk}
		}

\author{\speaker{Shigemi Ohta} 
		for RBC and UKQCD Collaborations\\
        Theory Center, Institute of Particle and Nuclear Studies, KEK, Tsukuba, Ibaraki, 3050801, Japan\\
        Department of Particle and Nuclear Physics, SOKENDAI, Hayama, Kanagawa, 2400193, Japan\\
        RIKEN BNL Research Center, BNL, Upton, NY 11973, USA\\
        E-mail: \email{shigemi.ohta@kek.jp}
        }

\abstract{
Nucleon isovector vector- and axialvector-current form factors, the renormalized isovector transversity and scalar charge, and the bare quark momentum and helicity moments of isovector structure functions  are reported with improved statistics from two recent RBC+UKQCD 2+1-flavor dynamical domain-wall fermions ensembles: Iwasaki\(\times\)DSDR gauge \(32^3\times64\) at inverse lattice spacing of 1.38 GeV and pion mass of 249 and 172 MeV.

\vspace{-189mm}\parbox{\textwidth}{\flushright\large\rm \hfill KEK-TH-1939, RBRC-1204}\vspace{182mm}
}

\FullConference{34th annual International Symposium on Lattice Field Theory\\
		24-30 July 2016\\
		University of Southampton, UK}

\begin{document}

\section{Introduction}

The  RIKEN-BNL-Columbia (RBC) Collaboration and the joint RBC and UKQCD Collaborations have been calculating some nucleon isovector observables \cite{Yamazaki:2008py,Lin:2008uz,Yamazaki:2009zq,Aoki:2010xg,Ohta:2010sr,Ohta:2011vv,Lin:2012nv,Ohta:2013qda,Lin:2014saa,Ohta:2014rfa} in numerical lattice QCD using ensembles generated with dynamical domain-wall fermion (DWF) quarks \cite{Aoki:2004ht,Allton:2008pn,Aoki:2010dy,Arthur:2012yc}.
Over the course we found the vector-current form factors behave well, and together with some moments of structure functions such as quark momentum and helicity fractions, appear trending to the respective experiments as the pion mass is set lower toward the experiment.
In contrast we discovered the axial charge is seriously underestimated on the lattice.
By this year's conference most calculations with various different actions at similar hadron mass and lattice cut off have confirmed this deficit \cite{Dragos:2016rtx,Bhattacharya:2016zcn,Liu:Lat2016}. 
Especially important for calculations with Wilson-fermion quarks is to remove the systematic errors at the linear order in the lattice spacing.
This systematics is exponentially suppressed in calculations with DWF quarks. 
More recently another joint collaboration of RBC and Lattice Hadron Physics (LHP) Collaborations was formed to calculate nucleon structure using RBC+UKQCD dynamical 2+1-flavor DWF ensembles at physical mass \cite{Blum:2014tka}.  
The first results from this joint collaboration were reported two years ago \cite{Syritsyn:2014xwa}.
Unfortunately the progress of this effort since then has been slow due to shortage of eligible computers \cite{Ohta:2015aos}.
This year I report new analysis of nucleon structure using two ensembles with Iwasaki\(\times\)DSDR gauge action with a lattice cut off of about 1.378(7) GeV and pion mass at about 249.4(3) and 172.3(3) MeV.

\section{Numerics}

We follow the conventional method of taking appropriate ratios of nucleon three-point to two-point functions, as described in our earlier publications \cite{Lin:2008uz,Yamazaki:2009zq,Aoki:2010xg}.
We take advantage of the exact isospin SU(2) symmetry of our degenerate up and down flavors.
Nucleon sources are optimized with gauge-invariant Gaussian smearing \cite{Alexandrou:1992ti,Berruto:2005hg} in order to reduce excited-state contamination.
We chose a Gaussian width of six lattice spacings for the present work \cite{Ohta:2010sr}.
By comparing results from source-sink separations of seven and nine lattice spacings, we had established absence of excited-state contamination by the former, shorter separation \cite{Ohta:2013qda,Lin:2014saa} in observables we are reporting.
Here I am reporting the results from the latter, longer separation with much bigger statistics.

We use the two recent joint RBC+UKQCD ensembles with Iwasaki\(\times\)DSDR gauge action at the gauge coupling of \(\beta=1.75\) \cite{Arthur:2012yc}.  
The inverse lattice spacing for these ensembles are now estimated at \(a^{-1} = 1.378(7)\) GeV \cite{Blum:2014tka}.
The strange-quark mass is set at 0.0045 lattice units, essentially at its physical value.
The degenerate up and down quark mass of 0.0042 and 0.001 in lattice units  respectively correspond to the pion mass, \(m_\pi\), of 249.4(3) and 172.3(3) MeV.
With the \(32^3\times64\) four-dimensional lattice volume, these ensembles are with \((L=4.58(2) {\rm fm})^3\) spatial volume and finite-size scaling parameter, \(m_\pi L\), of 5.79(6) and 4.00(6) respectively.
For the 249-MeV ensemble we use 165 configurations separated from each other by 8 molecular dynamics (MD) trajectories, each with seven source positions for conventional measurements.
For the 172-MeV ensemble we use the ``AMA'' technique \cite{Shintani:2014vja} with 39 configurations separated from each other by 16 MD trajectories, each with 112 sloppy and 4  accurate measurements.

\section{Results}

\begin{figure}[b]
\begin{center}
\includegraphics[width=0.8\textwidth,clip]{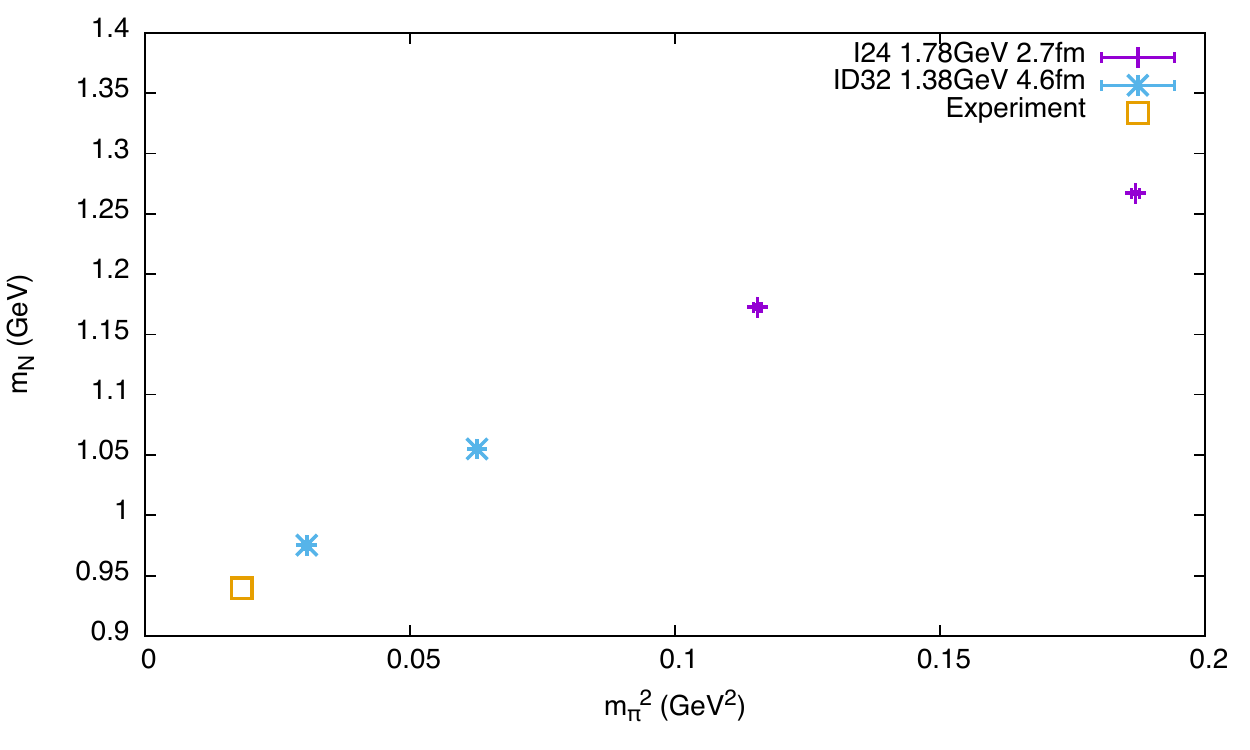}
\caption{\label{fig:mpi2mN}
Nucleon mass plotted against pion mass squared from the present Iwasaki\(\times\)DSDR ensembles at \(a^{-1}=1.378(7)\) GeV \cite{Arthur:2012yc} along with the two from earlier Iwasaki ensembles at \(a^{-1}=1.7848(5)\) GeV \cite{Allton:2008pn}.
}
\end{center}
\end{figure}

In Table \ref{tab:mN} and Figure \ref{fig:mpi2mN} our nucleon mass estimates are summarized, both for the present Iwasaki\(\times\)DSDR ensembles at \(a^{-1}=1.378(7)\) GeV \cite{Arthur:2012yc} and for the two lightest of earlier Iwasaki ensembles at \(a^{-1}=1.7848(5)\) GeV \cite{Allton:2008pn}.
The present, lighter-mass, results trend much better than the earlier heavier ones toward the experiment.
They capture the experimental point within the statistical error in linear extrapolation in the pion mass squared.
Note in these four ensembles we slightly revised the lattice scales recently \cite{Blum:2014tka}.
\begin{table}[t]
\begin{center}
\begin{tabular}{llll}
\hline
\multicolumn{1}{c}{\(a^{-1}\)[GeV]}&
\multicolumn{1}{c}{\(m_qa\)} &
\multicolumn{1}{c}{\(m_Na\)} &
\multicolumn{1}{c}{\(m_N\) [GeV]}\\
\hline\hline
1.378(7) & 0.001 & 0.7077(8) & 0.9752(11)\\
               & 0.0042 & 0.7656(2) & 1.0550(3)\\
\hline
1.7848(5)  & 0.005 & 0.6570(9)& 1.1726(16)\\
                 & 0.01    & 0.7099(5) & 1.2670(9)\\
\hline
\end{tabular}
\end{center}
\caption{\label{tab:mN}
Nucleon mass estimates for the present Iwasaki\(\times\)DSDR ensembles at \(a^{-1}=1.378(7)\) GeV \cite{Arthur:2012yc} along with those for two earlier Iwasaki ensembles at \(a^{-1}=1.7848(5)\) GeV \cite{Allton:2008pn}.
}
\end{table}

\begin{figure}[t]
\hfill
\includegraphics[width=0.45\textwidth,clip]{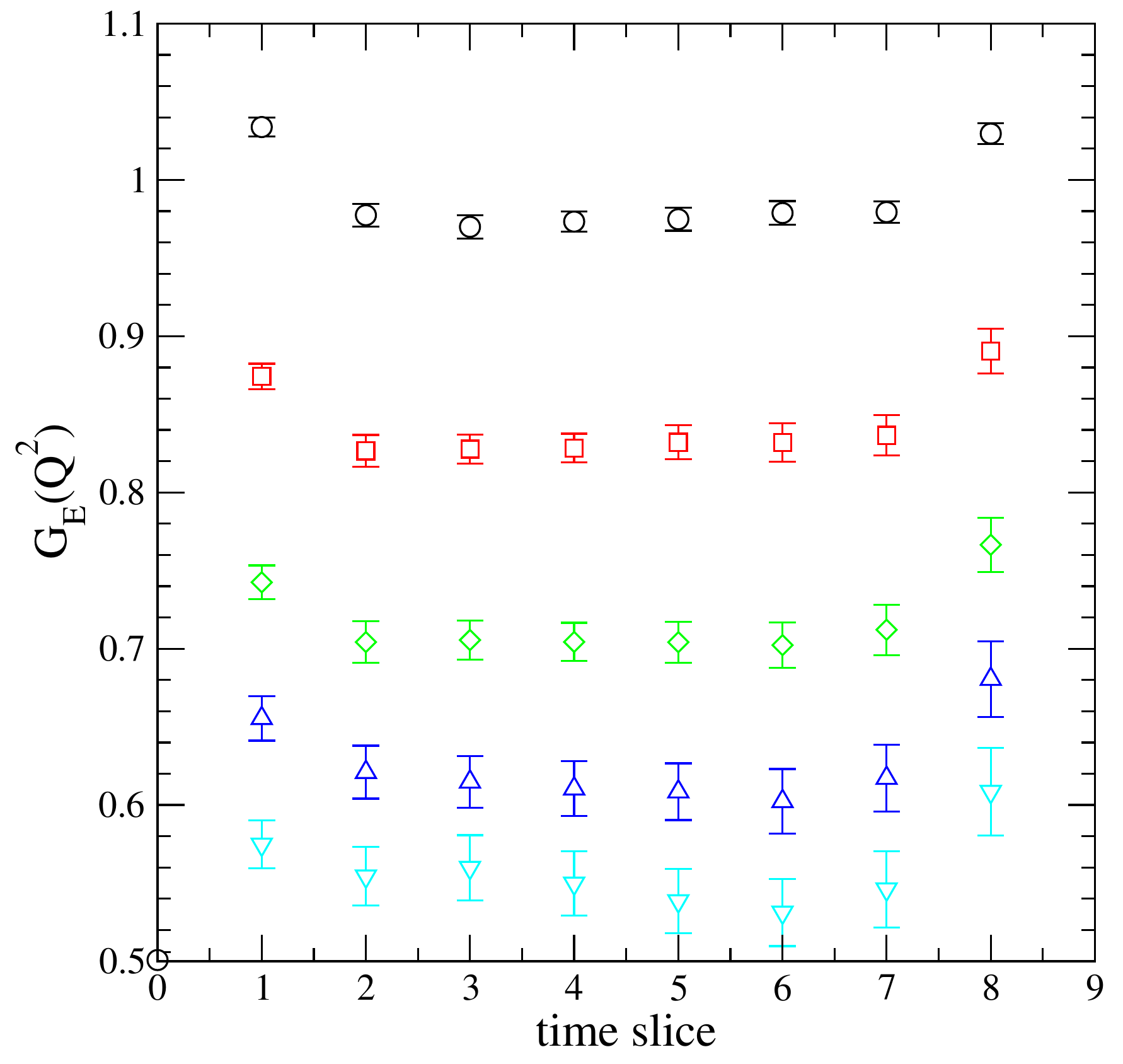}\hfill
\includegraphics[width=0.45\textwidth,clip]{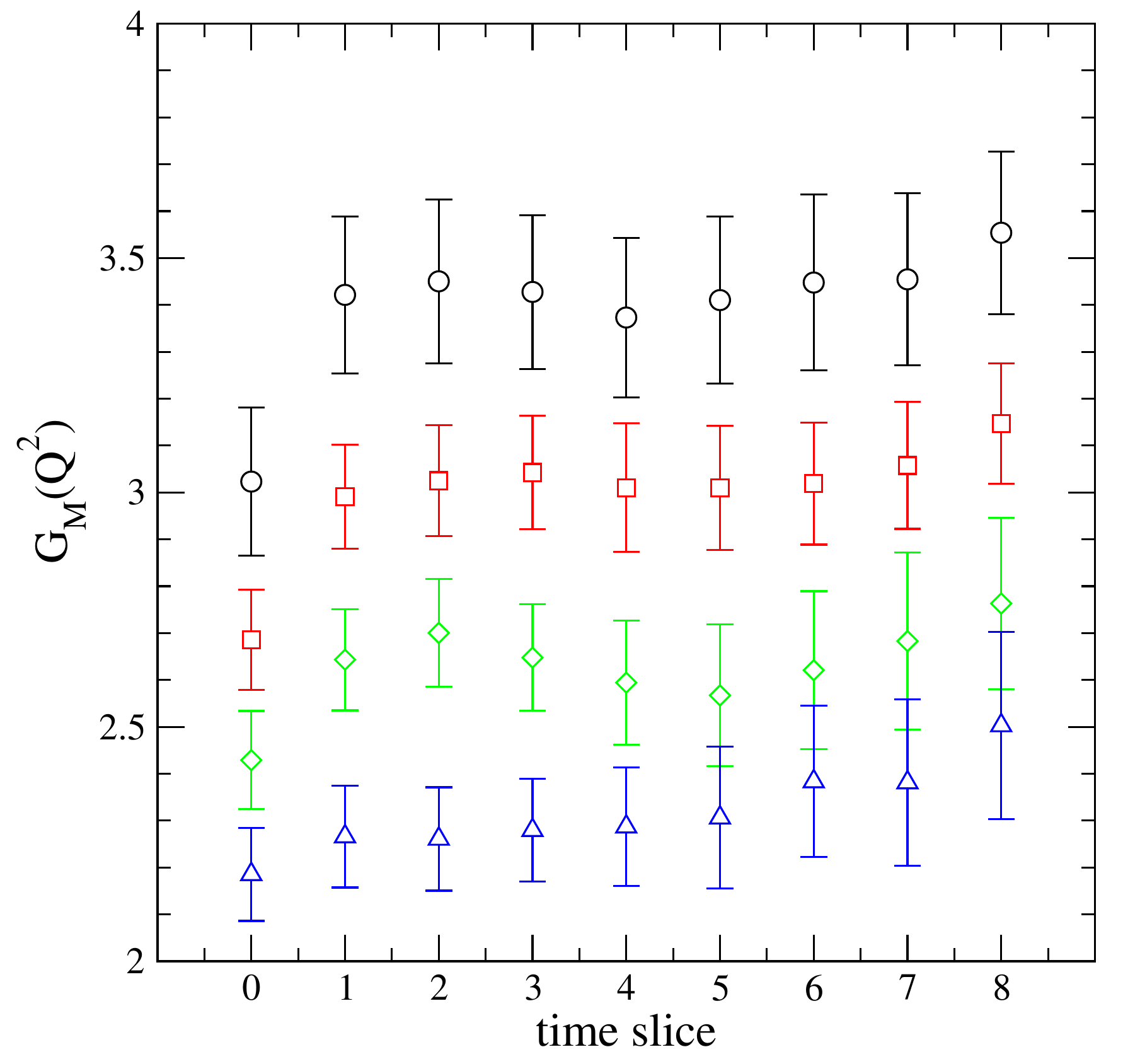}\hfill
\caption{\label{fig:EMFF}
Signal qualities for isovector vector electric, \(G_E(Q^2)\), and magnetic, \(G_M(Q^2)\), form factors for the 172-MeV ensemble.
Symbol colors, black, red, green, blue, and light blue respectively correspond to momentum transfer squared, \(Q^2\), of 0, 1, 2, 3, and 4 in lattice units.
}
\end{figure}

In Figure \ref{fig:EMFF} we present the signal quality of the isovector vector electric, \(G_E(Q^2)\), and magnetic, \(G_M(Q^2)\), form factors for the 172-MeV ensemble.
The quality is similar, or somewhat better, for the 249-MeV ensemble.
The results are consistent with what Meifeng Lin reported earlier \cite{Lin:2014saa}.
By fitting these with the conventional dipole form, we obtain the isovector vector Dirac, \(r_1\), and Pauli, \(r_2\), root-mean-squared radii as well as the isovector anomalous magnetic moment, \(F_2(0)\), presented in Table \ref{tab:DiracPauli}.
\begin{table}
\begin{center}
\begin{tabular}{llll}
\hline
\multicolumn{1}{c}{\(m_\pi\) (MeV)}& 
\multicolumn{1}{c}{\(r_1\) (fm)} &
\multicolumn{1}{c}{\(r_2\) (fm)} &
\multicolumn{1}{c}{\(F_2(0)\)} \\
\hline\hline
172 & 0.63(5) & 0.84(6)& 3.3(2) \\
249 & 0.62(11)& 0.77(9)& 3.2(3)\\
\hline
\end{tabular}
\end{center}
\caption{\label{tab:DiracPauli}
Conventional diopole fits to the isovector Dirac, \(F_1\), and Pauli, \(F_2\), form factors.
}
\end{table}
We use the measured nucleon mass in Table \ref{tab:mN} for normalizations of induced form factors such as \(F_2\).
We are in a process of moving our analysis away from the dipole form to a less model-dependent form such as \(z\)-expansion for these vector- as well as the axialvector-current form factors, but are not yet ready to quote numbers.

The ratio of the isovector axialvector, \(g_A\), and vector, \(g_V\), charges from these ensembles was reported earlier \cite{Ohta:2013qda}: \(g_A/g_V=1.15(4)\) and 1.17(4) respectively for the 172-MeV and 249-MeV ensembles.
These estimates need no revision at this time, and now appear to have been confirmed by a few other major collaborations \cite{Dragos:2016rtx,Bhattacharya:2016zcn,Liu:Lat2016} using different actions but with similar lattice spacings and quark masses.
Especially important for calculations with Wilson-fermion quarks \cite{Dragos:2016rtx,Bhattacharya:2016zcn} is to remove the \(O(a)\) systematic errors at the linear order in the lattice spacing, \(a\) \cite{Liu:Lat2016}. 
These systematics are exponentially suppressed in our DWF formulation.
We had also demonstrated the nucleon isovector observables we are calculating are not affected by excited-state contamination once we optimize our source and source-sink separation \cite{Ohta:2014rfa}.
Thus the finite lattice spatial volumes remain the least-studied systematics.

\begin{figure}[t]
\hfill
\includegraphics[width=0.45\textwidth,clip]{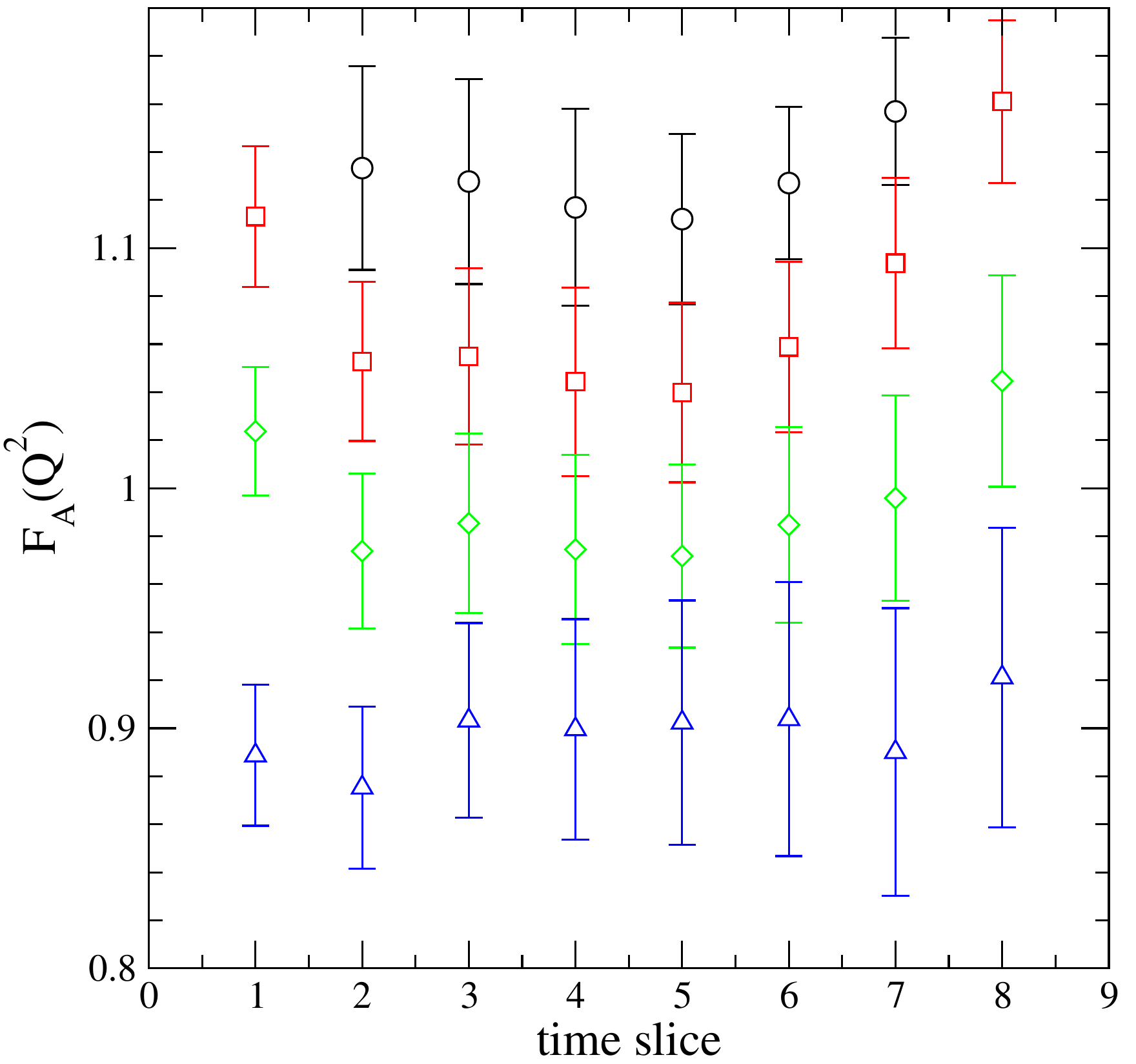}\hfill
\includegraphics[width=0.45\textwidth,clip]{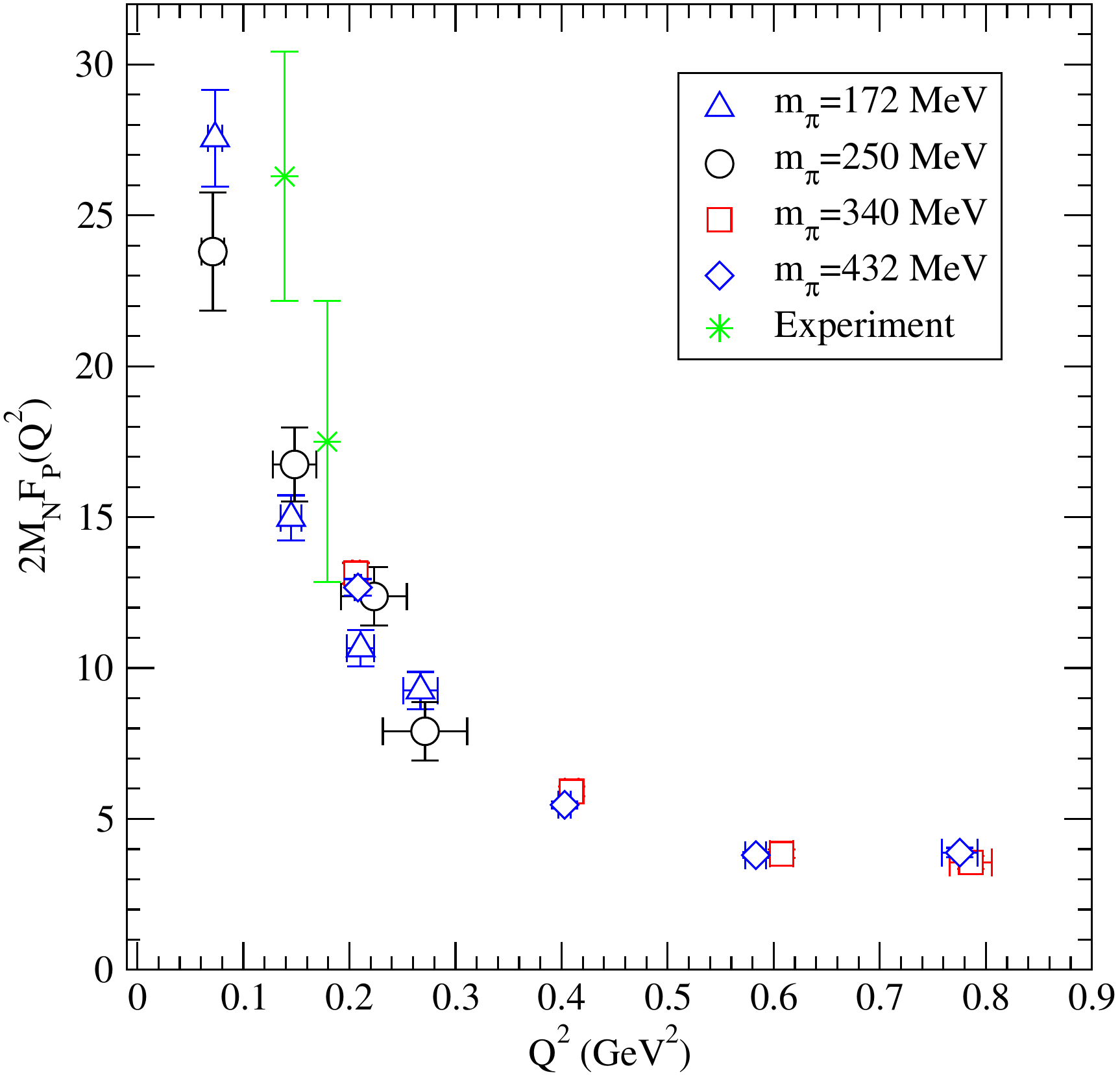}\hfill
\caption{\label{fig:FAFP}
Signal qualities for isovector axialvector-current form factors for the 172-MeV ensemble.
}
\end{figure}

\begin{figure}[b]
\begin{center}
\includegraphics[width=0.8\textwidth,clip]{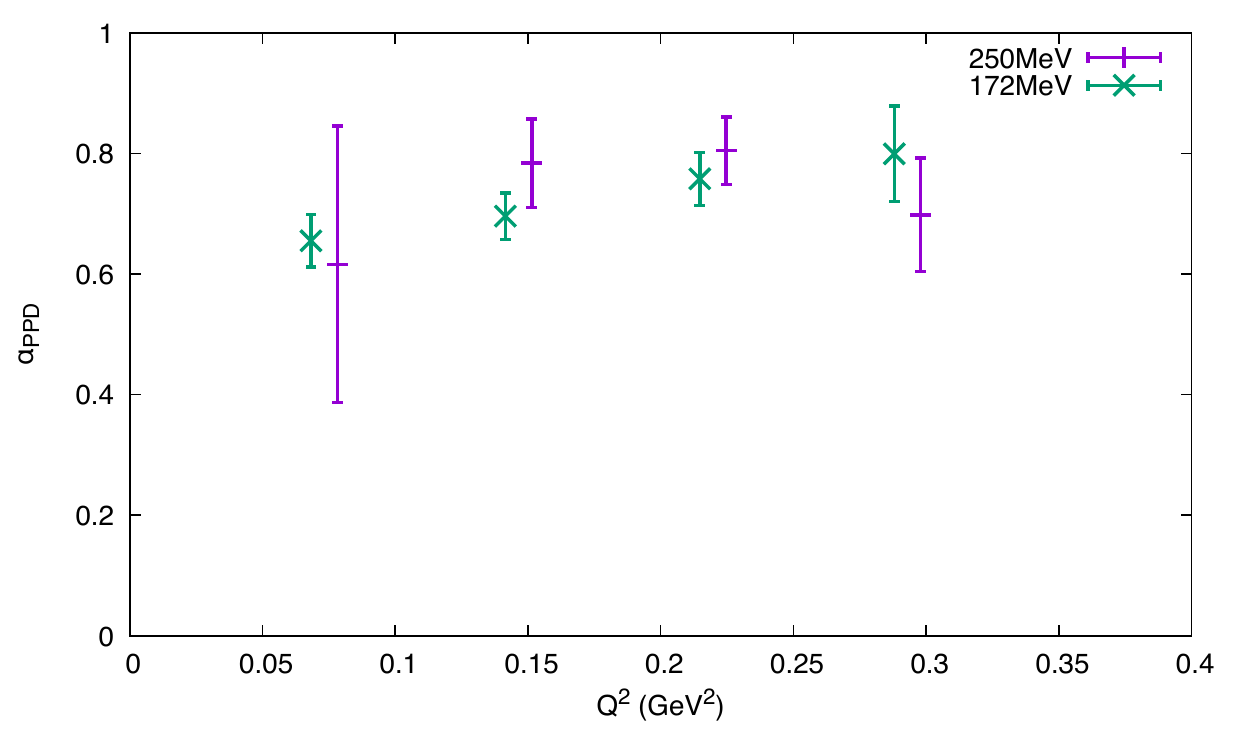}
\end{center}
\caption{\label{fig:PPD}
Pion-pole-dominance measure, \(\displaystyle \alpha_{PPD}=\frac{(m_\pi^2+q^2)F_P(q^2)}{2m_N F_A(q^2)}\).
}
\end{figure}

Signal quality for the isovector axialvector-current form factors are presented in Figure \ref{fig:FAFP}.
From the axialvector form factors we can extract its root-mean-squared radius as 0.55(6) and 0.54(6) fm respectively for the 172- and 249-MeV ensembles via the conventional dipole-form fitting.
From the pseudoscalar form factors we can test the pion-pole dominance, as is presented in Figure \ref{fig:PPD}.
It appears flat in both ensembles but away from the unity.
We likely need larger spatial volume to investigate this further.

\begin{figure}[t]
\includegraphics[width=0.49\textwidth]{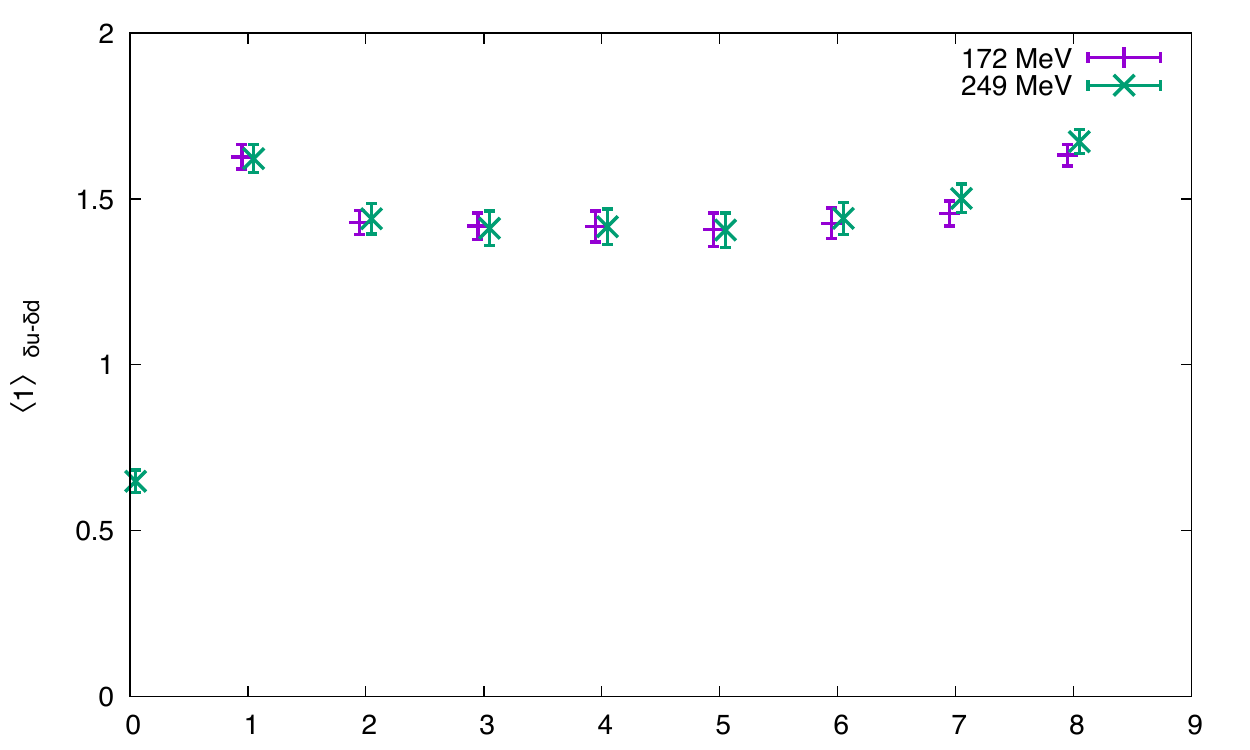}
\includegraphics[width=0.49\textwidth]{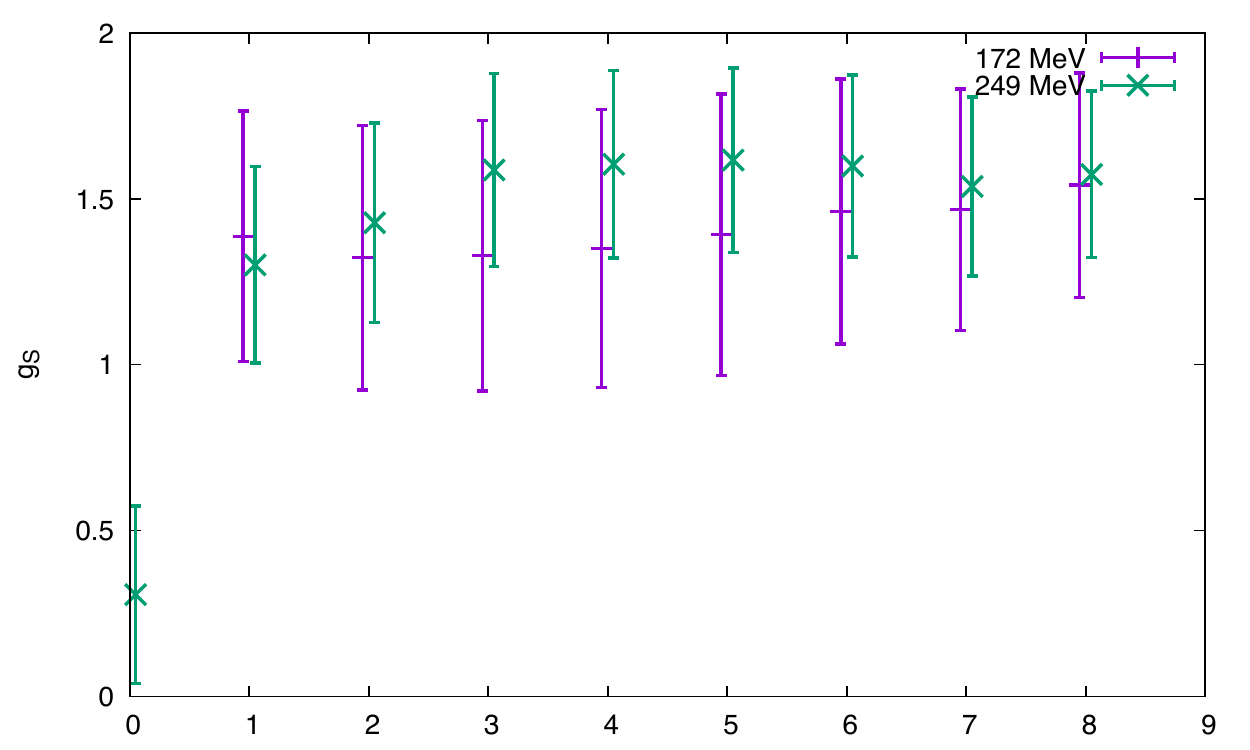}
\caption{\label{fig:1q}
Bare isovector transverstiy, \(\langle 1 \rangle_{\delta u - \delta d}\), and scalar charge, \(g_S\), plateau signals.
}
\end{figure}

Plateau signals for the bare isovector transversity, \(\langle 1 \rangle_{\delta u - \delta d}\), and the scalar charge, \(g_S\), are presented in Figure \ref{fig:1q}.
The transversity signals are very clean and do not show any mass dependence.
As was reported two years ago, the isovector tranversity shows weaker but still relevant signs of long-lasting autocorrelation similar to that of the axial charge in the lighter, 172-MeV, ensemble \cite{Ohta:2014rfa}.
Yet the agreement with the heavier ensemble where there is no such autocorrelation reassures this is less problematic here in the transversity than in the axial charge.
The scalar plateaus are also well defined albeit with larger statistical errors.
No mass dependence can be seen here either.
We have completed non-perturbative renormalizations for these observables:
\(
Z_S({\rm RI/SMOM}, \mu={\rm 2.0 GeV})=0.619(08)_{\rm stat}(24)_{\rm syst},
\)
and
\(
Z_T({\rm RI/SMOM}, \mu={\rm 2.0 GeV})=0.731(08)_{\rm stat}(29)_{\rm syst}.
\)
From these, we obtain our estimates for the renormalized isovector transversity and scalar charge as presented in Table \ref{tab:TS}.
\begin{table}
\begin{center}
\begin{tabular}{lll}
\hline
\multicolumn{1}{c}{\(m_\pi\) [MeV]} &
\multicolumn{1}{c}{\(\langle 1 \rangle_{\delta u-\delta d}\)} &
\multicolumn{1}{c}{\(g_S\)}\\
\hline\hline
172& \(1.42(4) \times 0.73(3) = 1.05(5)\) & \(1.4(4) \times 0.62(3) = 0.9(3)\) \\
249& \(1.42(5) \times 0.73(3) = 1.05(5)\) & \(1.6(3) \times 0.62(3) = 1.0(2)\) \\
\hline
\end{tabular}
\end{center}
\caption{\label{tab:TS}
Renormalized isovector transverstiy, \(\langle 1 \rangle_{\delta u - \delta d}\), and scalar charge, \(g_S\).
}
\end{table}
The transversity errors are dominated by a scheme-dependence systematics in non-perturbative renormalization, at about five percent, due mainly from the relatively low lattice cut off.
The scalar errors are still dominated by statistical noise, but will eventually encounter the same scheme-dependence systematics in non-perturbative renormalization.

\begin{figure}[b]
\includegraphics[width=0.49\textwidth,clip]{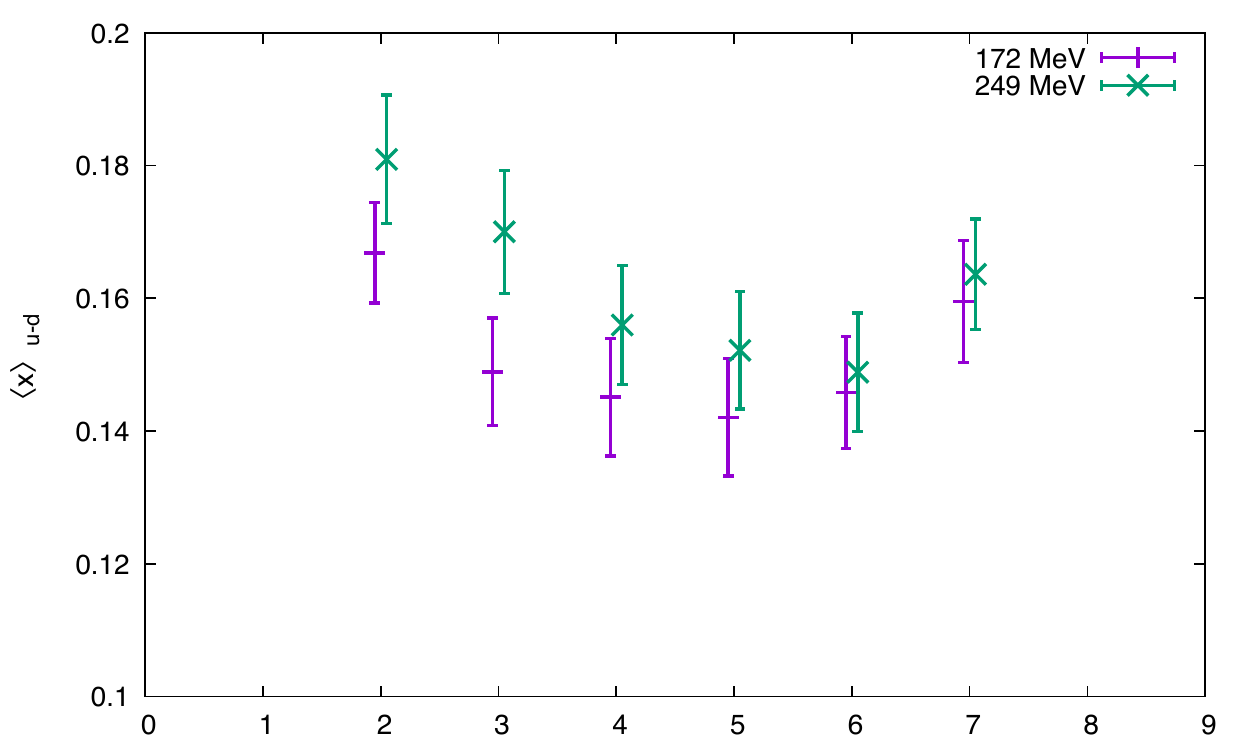}
\includegraphics[width=0.49\textwidth,clip]{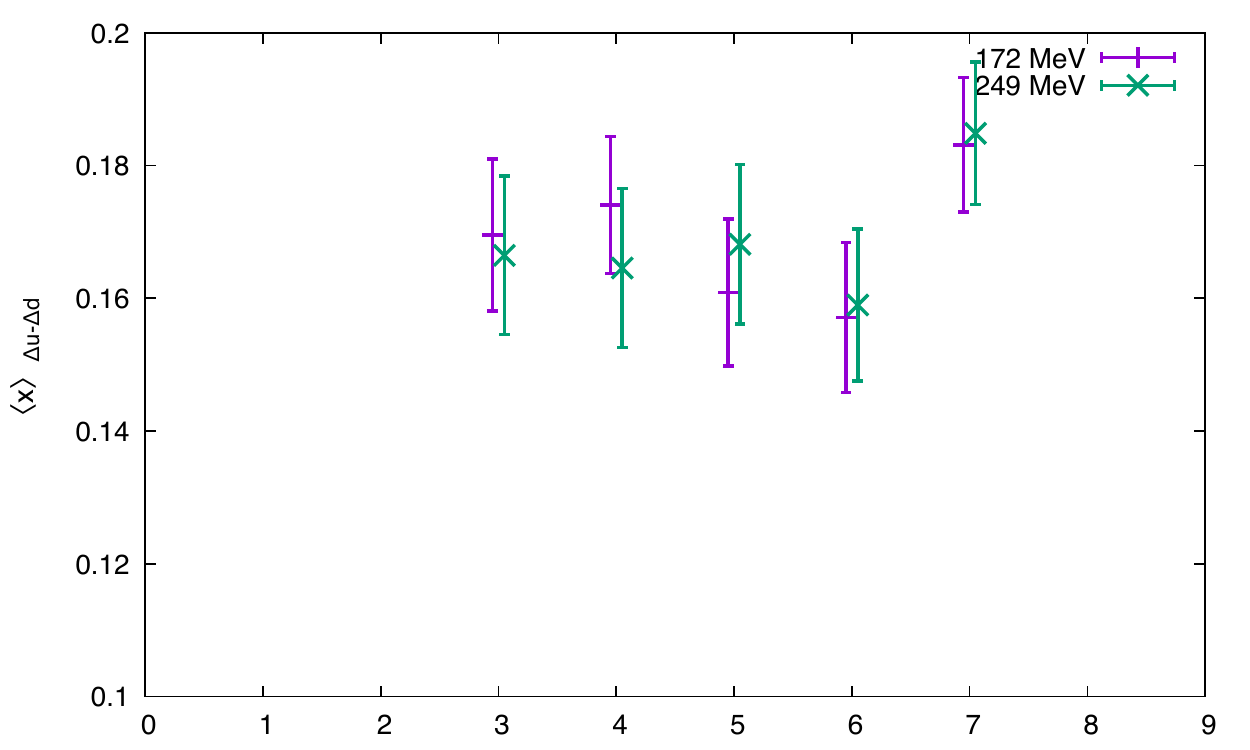}
\caption{\label{fig:fractions}
Plateau signals for the bare isovector quark momentum, \(\langle x \rangle_{u-d}\), and helicity, \(\langle x \rangle_{\Delta u-\Delta d}\).
}
\end{figure}

We present plateau signals for the bare isovector quark momentum, \(\langle x \rangle_{u-d}\), and helicity, \(\langle x \rangle_{\Delta u-\Delta d}\), fractions in Figure \ref{fig:fractions}.
As can be seen these signals are noisier than the form factors, transversity, and scalar charge: yet from fitting these in the range from 3 to 6 lattice units the bare isovector quark momentum and helicity fractions are obtained as in Table \ref{tab:fractions}.
\begin{table}
\begin{center}
\begin{tabular}{lll}
\hline
\multicolumn{1}{c}{\(m_\pi\) [MeV]} &
\multicolumn{1}{c}{\(\langle x \rangle_{u-d}\)} &
\multicolumn{1}{c}{\(\langle x \rangle_{\Delta u-\Delta d}\)}\\
\hline\hline
172& 0.145(7)& 0.165(9)\\
249& 0.157(7)& 0.165(10)\\
\hline
\end{tabular}
\end{center}
\caption{\label{tab:fractions}
Bare isovector quark momentum, \(\langle x \rangle_{u-d}\), and helicity, \(\langle x \rangle_{\Delta u-\Delta d}\).
}
\end{table}
While the momentum fraction may still be slowly decreasing with the mass, the helicity fraction appears to stay flat.
As we are yet to renormalize these, it is not possible to compare them with their counterparts from the earlier calculations at a finer lattice spacing and heavier masses \cite{Aoki:2010xg}.
However the trending down of these observables toward the experiments seen in the earlier calculations at heavier masses has at least slowed down and possibly stopped by the present mass ranges.

Signals for the twist-3, \(d_1\), moment of the isovector polarized structure function are even noisier than the momentum and helicity fractions and are yet to provide any finite value.

In summary we are finalizing our analysis of nucleon isovector form factors, transversity, scalar charge, and quark momentum and helicity fractions for the two recent RBC+UKQCD dynamical 2+1-flavor DWF ensembles with pion mass of 249.4(3) and 172.3(3) MeV.


\begin{thebibliography}{99}

\bibitem{Yamazaki:2008py} 
  T.~Yamazaki {\it et al.} [RBC and UKQCD], 
  Phys.\ Rev.\ Lett.\  {\bf 100}, 171602 (2008)
  [arXiv:0801.4016].

\bibitem{Lin:2008uz} 
  H.~W.~Lin {\it et al.} [RBC], 
  Phys.\ Rev.\ D {\bf 78}, 014505 (2008)
  [arXiv:0802.0863].

\bibitem{Yamazaki:2009zq} 
  T.~Yamazaki {\it et al.} [RBC], 
  Phys.\ Rev.\ D {\bf 79}, 114505 (2009)
  [arXiv:0904.2039].

\bibitem{Aoki:2010xg} 
  Y.~Aoki {\it et al.} [RBC], 
  Phys.\ Rev.\ D {\bf 82}, 014501 (2010)
  [arXiv:1003.3387].

\bibitem{Ohta:2010sr} 
  S.~Ohta [RBC and UKQCD], 
  PoS LATTICE {\bf 2010}, 152 (2010)
  [arXiv:1011.1388].

\bibitem{Ohta:2011vv} 
  S.~Ohta [RBC and UKQCD], 
  PoS LATTICE {\bf 2011}, 168 (2011)
  [arXiv:1111.5269].

\bibitem{Lin:2012nv} 
  M.~Lin {\it et al.} [RBC and UKQCD], 
  PoS LATTICE {\bf 2012}, 171 (2012)
  [arXiv:1212.3235].

\bibitem{Ohta:2013qda} 
  S.~Ohta [RBC and UKQCD], 
  PoS LATTICE {\bf 2013}, 274 (2014)
  [arXiv:1309.7942].

\bibitem{Lin:2014saa} 
  M.~Lin,
  PoS LATTICE {\bf 2013}, 275 (2014)
  [arXiv:1401.1476 [hep-lat]].

\bibitem{Ohta:2014rfa} 
  S.~Ohta [RBC and UKQCD], 
  PoS LATTICE {\bf 2014}, 149 (2014)
  [arXiv:1410.8353].
   
\bibitem{Aoki:2004ht} 
  Y.~Aoki {\it et al.} [RBC],
  Phys.\ Rev.\ D {\bf 72}, 114505 (2005)
  [hep-lat/0411006].

\bibitem{Allton:2008pn} 
  C.~Allton {\it et al.} [RBC and UKQCD], 
  Phys.\ Rev.\ D {\bf 78}, 114509 (2008)
  [arXiv:0804.0473].

\bibitem{Aoki:2010dy} 
  Y.~Aoki {\it et al.} [RBC and UKQCD], 
  Phys.\ Rev.\ D {\bf 83}, 074508 (2011)
  [arXiv:1011.0892].
  
\bibitem{Arthur:2012yc} 
  R.~Arthur {\it et al.} [RBC and UKQCD], 
  Phys.\ Rev.\ D {\bf 87}, 094514 (2013)
  [arXiv:1208.4412].

\bibitem{Dragos:2016rtx} 
  J.~Dragos, R.~Horsley, W.~Kamleh, D.~Leinweber, Y.~Nakamura, P.~Rakow, G.~Schierholz, R.~Young, and J.~Zanotti,
  [arXiv:1606.03195].

\bibitem{Bhattacharya:2016zcn} 
  T.~Bhattacharya, V.~Cirigliano, S.~Cohen, R.~Gupta, H.~W.~Lin and B.~Yoon,
  Phys.\ Rev.\ D {\bf 94}, 
  054508 (2016)
  [arXiv:1606.07049].
  
\bibitem{Liu:Lat2016}
  J.~Liang, Y.~Yang, K.-F.~Liu, A.~Alexandru, T.~Draper, and R. S.~Sufian, 
  in these proceedings.

\bibitem{Blum:2014tka} 
  T.~Blum {\it et al.} [RBC and UKQCD], 
  Phys.\ Rev.\ D {\bf 93}, 074505 (2016)
  [arXiv:1411.7017].
  
\bibitem{Syritsyn:2014xwa} 
  S.~Syritsyn {\it et al.} [LHP and RBC and UKQCD],
  PoS LATTICE {\bf 2014}, 134 (2015)
  [arXiv:1412.3175].
 
\bibitem{Ohta:2015aos} 
  S.~Ohta [LHP and RBC and UKQCD], 
  PoS LATTICE {\bf 2015}, 124 (2016)
  [arXiv:1511.05126].
  
\bibitem{Alexandrou:1992ti} 
  C.~Alexandrou, S.~Gusken, F.~Jegerlehner, K.~Schilling and R.~Sommer,
  Nucl.\ Phys.\ B {\bf 414}, 815 (1994)
  [hep-lat/9211042].
  
\bibitem{Berruto:2005hg} 
  F.~Berruto, T.~Blum, K.~Orginos and A.~Soni,
  Phys.\ Rev.\ D {\bf 73}, 054509 (2006)
  [hep-lat/0512004].

\bibitem{Shintani:2014vja} 
  E.~Shintani, R.~Arthur, T.~Blum, T.~Izubuchi, C.~Jung and C.~Lehner,
  Phys.\ Rev.\ D {\bf 91}, 
  114511 (2015)
  [arXiv:1402.0244].

\end{thebibliography}
\end{document}